\documentclass[twocolumn,preprintnumbers]{revtex4}

\usepackage{epsfig,psfrag}
\usepackage{amsmath,amssymb}
\usepackage{bm}
\usepackage{hyperref}
\usepackage{epsfig}
\usepackage{epic}
\usepackage{mathrsfs}

\newcommand{\BesselJ}{{J}}

\newcommand{\sech}{\mathrm{sech}}

\newcommand{\Cusp}{\Gamma_{\mathrm{cusp}}}

\newcommand{\Op}{\mathcal{O}}

\newcommand{\alg}[1]{\mathfrak{#1}}
\newcommand{\superN}{\mathcal{N}}

\newcommand{\EulerGamma}{\gamma_{\indups{E}}}
\newcommand{\indups}[1]{_{\mathrm{\scriptscriptstyle #1}}}

\makeatletter
\def\mr@ignsp#1 {\ifx\:#1\@empty\else #1\expandafter\mr@ignsp\fi}%
\newcommand{\multiref}[1]{\begingroup
\xdef\mr@no@sparg{\expandafter\mr@ignsp#1 \: }%
\def\mr@comma{}%
\@for\mr@refs:=\mr@no@sparg\do{\mr@comma\def\mr@comma{,}\ref{\mr@refs}}%
\endgroup}
\makeatother
\long\def\symbolfootnote[#1]#2{\begingroup%
\def\thefootnote{\fnsymbol{footnote}}\footnote[#1]{#2}\endgroup}

\begin{document}

\preprint{AEI-2009-005}
\preprint{NSF-KITP-09-06}
\preprint{UUITP-02/09}
\title{The virtual scaling function of AdS/CFT}
\author{Lisa Freyhult$^a$ and Stefan Zieme$^{b,c}$}
\affiliation{$^a$ Department of Physics and Astronomy, Uppsala University, SE-751 08 Uppsala, Sweden}
\affiliation{$^b$ Max-Planck-Institut f\"ur Gravitationsphysik, Albert-Einstein-Institut, 
    Am M\"uhlenberg 1, 14476 Golm, Germany}
\affiliation{$^c$Kavli Institute for Theoretical Physics, University of California Santa Barbara, CA 93106 USA}
\begin{abstract}
We write an integral equation that
incorporates finite corrections to the large spin asymptotics 
of $\superN=4$ SYM twist operators from the non-linear integral equation. We show that these 
corrections are an all-loop result, not affected by wrapping effects, and agree, after 
determining the strong coupling expansion, with string theory predictions. 
\end{abstract}
\maketitle
\section{Introduction}
Twist operators have proved to be very important operators for studying the planar limit of 
the AdS/CFT correspondence.Their special scaling behavior at large spin \cite{Belitsky:2006en} 
was established and an integral equation describing the logarithmic growth of the anomalous dimension 
was derived  \cite{Beisert:2006ez}. In the weak coupling regime its expansion agrees with the explicit 
four-loop $\superN=4$ result \cite{Bern:2006ew}. 

The strong coupling limit was studied in \cite{Basso:2007wd,Kostov:2008ax} and shown to
match the available two-loop string data \cite{Cusp string results} while predictions for higher loops were 
also made in \cite{Basso:2007wd}.

In particular, the anomalous dimension of twist $L$ operators exhibits the large spin $M$ scaling behavior
\begin{equation}\label{scaling}
 \gamma_L(g,M)=f(g) (\log M+\EulerGamma-(L-2)\log 2 )+B_L(g) + \ldots \,.
\end{equation}
The scaling function $f(g)$ is commonly denoted the cusp anomalous dimension $f(g)=2 \Cusp(g)$ as it 
describes the logarithmic growth of the anomalous dimension of light-like Wilson loops with a cusp \cite{Korchemsky:1985xj}. In analogy 
with the QCD splitting functions \cite{Moch:2004pa} we denote the finite order correction $B_L(g)$ as the 
{\it virtual} part.

In a refined limit for the scaling function an all-loop string 
calculation from the bosonic $O(6)$ sigma model has been performed \cite{Alday:2007mf}. It agrees to two loops
with the direct string computation \cite{Roiban:2007ju}. The same 
limit in the weak coupling regime leads to a generalized scaling function \cite{Freyhult:2007pz}
which agrees with the predictions from the $O(6)$ sigma model \cite{Alday:2007mf} upon continuation 
to infinite coupling, as was demonstrated in \cite{Basso:2008tx} and further studied in \cite{FRSstrongcoupling}. 
For an alternative approach to the generalized scaling function, see \cite{Bombardelli:2008ah}.

Anomalous dimensions of twist operators can also be computed for finite values of the spin in terms 
of harmonic sums \cite{Kotikov:2008pv}. For the leading twist-two operators the asymptotic Bethe ansatz 
and Baxter equation, however, break down at four-loop order \cite{Kotikov:2007cy} and wrapping effects 
have to be taken into account. 

For the lowest $M=2$ state, the Konishi operator, the complete anomalous dimension including wrapping effects have been
successfully computed from different perspectives \cite{Konishi}. Subsequently, using the L\"uscher formalism
the wrapping effects for twist-two operators, which cure the asymptotic Bethe ansatz result, 
were computed \cite{Bajnok:2008qj}. 
The result satisfies all BFKL \cite{Kotikov:2002ab} constraints at negative spins and predicts 
that at large values of the spin $M$ the first contributions from wrapping effects will be of order $\Op(\log^2 M/M^2 )$ \cite{Beccaria:2009vt}. 
This indicates that finite order corrections $\Op(M^0)$ to the large spin scaling can also be 
computed from the asymptotic integrable structure to all-loop orders in the Yang-Mills coupling.

In what follows we will compute these finite order effects to the logarithmic scaling behavior for operators of 
general twist $L$.  In the strong coupling regime we can rewrite the finite order corrections in terms of 
the functions that determine $\Cusp$ and recover the string results \cite{Beccaria:2008tg}, which shows
that the virtual corrections, $B_L(g)$, are indeed not affected by wrapping effects. We also determine subleading 
corrections in $1/g$. For twist-two operators we can also compute further subleading corrections in $M$ up to wrapping
order. They equally match string theory predictions. 

The cusp anomalous dimension describes the leading singularities in the logarithm of planar multi-loop gluon scattering 
amplitudes. The virtual part enters the subleading divergencies as a part 
of the collinear anomalous dimension \cite{Dixon:2008gr}. The latter are known to fourth order \cite{Bern:2005iz}.
%
\section{Finite order integral equation}
From the NLIE, see \cite{Feverati:2006tg} and references therein, of the $\alg{sl}(2)$ sector \cite{Freyhult:2007pz} one can 
extract the following equation for the density 
for the distribution of Bethe roots including corrections of $\Op(M^0)$
\begin{eqnarray}\label{eq: complete density}
\lefteqn{\hat{\sigma}(t)=\frac{t}{e^t-1} \Big[ K(2gt,0)(\log M +\EulerGamma-(L-2)\log2)} \nonumber\\
 &&-\frac{L}{8g^2 t}(\BesselJ_0(2gt) - 1) + \frac{1}{2}\int_{0}^{\infty} dt' \Big( \frac{2}{e^{t'}-1}-\frac{L-2}{e^{t'/2}+1}\Big)\nonumber\\
 &&\phantom{-\frac{L}{8g^2 t}(\BesselJ_0(2gt)} \times \left(K(2gt,2gt')-K(2gt,0)\right)  \nonumber\\
&&\phantom{\frac{L}{8g^2 t}(\BesselJ_0(2gt)}- 4g^2 \int_{0}^{\infty} dt' K(2gt,2gt')\hat{\sigma}(t') \Big] \,,
\end{eqnarray}
with the anomalous dimension given by $\gamma_L(g,M)=16g^2\,\hat{\sigma}(0)$.
The integral kernel $K(t,t')=K_0(t,t')+K_1(t,t')+K_d(t,t')$ can be written in terms
of Bessel function \cite{Beisert:2006ez}. The parity even/odd components are given by 
\begin{equation}
 K_0(t,t')=\frac{2}{t t'} \sum_{n=1}^{\infty} (2n-1)\BesselJ_{2n-1}(t) \BesselJ_{2n-1}(t')\,,\nonumber
 \end{equation}
\begin{equation}
 K_1(t,t')=\frac{2}{t t'} \sum_{n=1}^{\infty} (2n)\BesselJ_{2n}(t)\BesselJ_{2n}(t')\,,
\end{equation}
respectively. The dressing kernel $K_d(t,t')$ is given by the convolution
\begin{equation}
 K_d(t,t)=8g^2 \int_{0}^{\infty}dt'' K_1(t,2gt'')\frac{t''}{e^{t''}-1}K_0(2gt'',t')\,.
\end{equation}

We will drop the part $\sim(\log M + \EulerGamma-(L-2)\log2)$, readily identifiable as the contribution to the cusp 
anomalous dimension analyzed in \cite{Basso:2007wd}.
Decomposing the density into parity even/odd parts 
\begin{equation}
 \frac{e^t -1}{t}\sigma(t)=\frac{\gamma_+(2gt)}{2gt}+\frac{\gamma_-(2gt)}{2gt} \,,
\end{equation}
one can express the functions $\gamma_\pm(t)$ in the form of Neumann series over Bessel functions
\begin{eqnarray}\label{gamma_pm}
 \gamma_+ (t)&=& 2 \sum_{n=1}^{\infty}(2n) \BesselJ_{2n}(t) \gamma_{2n} \,, \nonumber\\
 \gamma_- (t)&=& 2 \sum_{n=1}^{\infty}(2n-1) \BesselJ_{2n-1}(t) \gamma_{2n-1} \,,
\end{eqnarray}
which satisfy an infinite system of equations. We want to obtain an equation for $B_L(g)$ in terms of 
the solution to the BES equation only, similar to the case of the generalized scaling function analyzed in 
\cite{Basso:2008tx}. Therefore, we introduce a label $j$ such that the system of equations becomes 
 \begin{eqnarray}\label{eq:even/odd j-n-system }
\int_{0}^{\infty} \frac{dt}{t}  
	           \Big[\frac{\gamma_+(t,j)}{1-e^{-t/2g}} &-& \frac{\gamma_-(t,j)}{e^{t/2g} - 1} \Big]\BesselJ_{2n}(t)
			=\frac{j L}{8ng}+j h_{2n}\,, \nonumber \\
 \int_{0}^{\infty} \frac{dt}{t}  
	           \Big[\frac{\gamma_-(t,j)}{1-e^{-t/2g}} &+& \frac{\gamma_+(t,j)}{e^{t/2g} - 1} \Big]\BesselJ_{2n-1}(t)
			=\frac{1-j}{2}\delta_{n,1} \nonumber \\
&&\hspace{3cm} + j h_{2n-1}\,,
\end{eqnarray}
with $h_n=h_n(g)$ given by 
\begin{equation}\label{eq:h_n}
 h_n(g)=\int_{0}^{\infty} \frac{dt}{4}  \Big(\frac{2}{e^{t}-1}-\frac{L-2}{e^{t/2}+1} \Big) 
			 	 	\Big( \frac{\BesselJ_{n}(2gt)}{gt}-\delta_{n,1}\Big).
\end{equation}
For $j=0$ one recovers the solution of the BES equation analyzed in \cite{Basso:2007wd}, while 
$j=1$ leads to the system of equations that determines $B_L(g)$.
To derive this set of equations we made use of the summation formula of even Bessel functions
$2\sum{}_{n=1}^{\infty} \BesselJ_{2n}(t)=(1-\BesselJ_0(t))$ and
the orthogonality relation of even/odd Bessel functions. 

As was shown in detail in \cite{Basso:2008tx}, we choose some reference $j'$ and multiply
both sides of the system \eqref{eq:even/odd j-n-system } with $(2n)\gamma_{2n}(g,j')$ and 
$(2n-1)\gamma_{2n-1}(g,j')$, respectively. Summing over all $n\ge 1$ and making use of the expansion
formulas of the even/odd parts \eqref{gamma_pm}  we obtain two equations for the even/odd parts 
$\gamma_\pm(t,j')$. Subtracting the even from the odd part one obtains an integral kernel which is 
invariant under $j \leftrightarrow j'$ as such should be its solution. This property, supplemented 
with the 
explicit form of $h_n$ \eqref{eq:h_n}, can be used to obtain $\gamma_1(g,j)$ in 
terms of the solution to the BES equation, $\gamma^{(0)}_\pm(t) \equiv \gamma_\pm(t,0)$ and 
$\gamma^{(0)}_1(t) \equiv \gamma_1(t,0)=f(g)/(16g^2)$, by 
putting $j'=0$ and taking $j=1$. Hence we obtain
\begin{eqnarray}
 &&\gamma_1(g,1)=
\frac{1}{4}\int_{0}^{\infty}dt \big[\frac{2}{e^t-1} - \frac{L-2}{e^{t/2}+1} \big] \times\\
	&&\big[\frac{\gamma^{(0)}_-(2gt)-\gamma^{(0)}_+(2gt)}{gt}-2\gamma_1(g,0) \big]
- \frac{L}{2g}\sum_{n=1}^{\infty}\gamma_{2n}(g,0).\nonumber
\end{eqnarray}
With the orthogonality of Bessel functions 
we obtain from \eqref{gamma_pm} 
\begin{equation}
 \gamma_{2n}=\int_{0}^{\infty} \frac{dt}{t} \BesselJ_{2n}(t)\gamma_+(t)\,,
\end{equation}
which 
can be used to derive for $B_L(g)=16g^2 \gamma_1(g,1)$
the final equation
\begin{eqnarray}
B_L(g)&=&4g^2 \int_{0}^{\infty}dt \big[\frac{2}{e^t-1}-\frac{L-2}{e^{t/2}+1} \big] \nonumber\\
&&\times \big[\frac{\gamma^{(0)}_-(2gt)-\gamma^{(0)}_+(2gt)}{gt}-2\gamma^{(0)}_1(g) \big]\nonumber\\
&&-4gL \int_{0}^{\infty} \frac{dt}{t} \gamma^{(0)}_+(2gt). 
\end{eqnarray}
At weak coupling one finds with the solution to the BES equation 
\begin{eqnarray}
 \gamma_-(2gt)&=&(1-g^2 \frac{\pi^2}{3})\BesselJ_1(2gt)+\Op(g^4) \,,\nonumber\\
 \gamma_+(2gt)&=&4g^3 \zeta_3 \BesselJ_2(2gt) + \Op(g^5) \,,
\end{eqnarray}
that $B_L(g)$, in agreement with \cite{Freyhult:2007pz}, has the expansion
\begin{equation}
 B_L(g)=-8 g^4 (7 - 2 L) \zeta_3 + \Op(g^6)\,.
\end{equation} 
%
\section{Strong coupling expansion}
In order to find the strong coupling expansion, according to \cite{Basso:2007wd},
we rewrite $\gamma_1(g,1) \equiv \gamma_1^{(1)}(g)$  as 
\begin{eqnarray}\label{gamma1}
\gamma_1^{(1)}(g)&=&\frac{1}{2}\int_0^\infty dt 
\Big( \frac{\gamma_-^{(0)}(t)}{g\,t\,(e^{t/2g}-1)}+\frac{\gamma_+^{(0)}(t)}{g\,t\,(e^{-t/2g}-1)} \nonumber\\
&-&\frac{\gamma_1^{(0)}(g)}{g(e^{t/2g}-1)} \Big) -\frac{(L-2)}{4}\int_0^\infty  dt \Big( \frac{\gamma_-^{(0)}(t)}{g\,t\,(e^{t/4g}+1)}
\nonumber\\
&+&\frac{\gamma_+^{(0)}(t)}{g\,t\,(e^{-t/4g}+1)}-\frac{\gamma_1^{(0)}(g)}{g(e^{t/4g}+1)}\Big) .
\end{eqnarray}
Following \cite{Basso:2007wd} we introduce the change of variables
\begin{equation}
2\gamma_\pm(t)= \big(1-\sech(t/2g)\big)\Gamma_\pm(t)\pm\tanh (t/2g) \Gamma_\mp(t)\,,
\end{equation}
and obtain $\gamma_1^{(1)}(g)$ as a function of the first generalized scaling function $\epsilon_1$ and 
the solution to the BES equation only 
\begin{eqnarray}\label{gamma1 Kvariable}
&&\gamma_1^{(1)}(g)=\frac{1}{16g^2}(L-2)\epsilon_1(g)+\gamma_1^{(0)}(g) (L-2) \log 2 \\
&&\phantom{\gamma_1}-\int_0^\infty dt\Big(\frac{1}{4g\,t}\big(\Gamma_+^{(0)}(t)+\Gamma_-^{(0)}(t)\big)+\frac{\gamma_1^{(0)}}{2g(e^{t/2g}-1)}\Big)\nonumber,
\end{eqnarray}
where $\epsilon_1$ is defined as in \cite{Basso:2008tx}. With the expansion of $\Gamma_{\pm}$
\begin{eqnarray}
\Gamma_+(t)&=&\sum_{k = 0}^\infty (-1)^{(k+1)} J_{2k}(t)  \Gamma_{2k} \,,\nonumber \\
\Gamma_-(t)&=&\sum_{k = 0}^\infty (-1)^{(k+1)} J_{2k-1}(t)\Gamma_{2k-1}\,,
\end{eqnarray}
according to \cite{Basso:2007wd} we have $\Gamma_0=4g\gamma_1^{(0)}$ and for $(k\ge-1)$
\begin{equation}
\Gamma_k=-\frac{1}{2}\Gamma_k^{(0)}+\sum_{p=1}^\infty \frac{1}{g^p}\left(c_p^-\Gamma_k^{(2p-1)}+c_p^+\Gamma_k^{(2p)}\right) ,
\end{equation}
with the coefficients $c_p^\pm$ given by $c_p^\pm=\sum_{r \geq 0} g^{-r}c_{p,r}^\pm$ and
\begin{equation}
\Gamma_{2m}^{(p)}=\frac{\Gamma(m+p-\frac{1}{2})}{\Gamma(m+1)\Gamma(\frac{1}{2})}, 
\quad \Gamma_{2m-1}^{(p)}=\frac{(-1)^p\Gamma(m-\frac{1}{2})}{\Gamma(m+1-p)\Gamma(\frac{1}{2})}\,.
\end{equation}
The part proportional to $\gamma_1^{(0)}$ of \eqref{gamma1 Kvariable} and the 
integral over Bessel functions can be performed to obtain 
\begin{eqnarray}\label{gamma1 final}
\gamma_1^{(1)}(g)&=&\frac{1}{16g^2}(L-2)\epsilon_1(g)+\gamma_1^{(0)}(g) (L-2) \log 2 \nonumber\\
&+&\gamma_1^{(0)} ( - \EulerGamma - \log g ) - \frac{1}{4g} \sum_{k=1}^{\infty}\frac{(-1)^{k+1}}{2k}\Gamma_{2k} \nonumber\\
&-& \frac{1}{4g} \sum_{k=0}^{\infty}\frac{(-1)^{k+1}}{2k-1}\Gamma_{2k-1}\,.
\end{eqnarray}
At strong coupling the first generalized scaling function has been analyzed in \cite{Basso:2008tx,Fioravanti:2008rv}
and is given by
\begin{equation}
 \epsilon_1(g)=-1+\Op(e^{-\pi g})\,.
\end{equation}
It does not receive perturbative $1/g$ corrections, but non-trivial exponential correction in the
coupling which are related to the mass gap parameter of the $O(6)$ model \cite{Basso:2008tx,Alday:2007mf}.
An all-order quantization condition that determines the coefficients $c_p^{\pm}$ has
been computed in \cite{Basso:2007wd}. Knowing these coefficients to a certain order 
readily gives $B_L(g)$ to the same order. To the leading order in $1/g$ they are given by
\begin{equation}
 c_1^+= - \frac{3 \log 2}{\pi} + \frac{1}{2}+ \Op(1/g) , \quad
 c_1^-=   \frac{3 \log 2}{4\pi} - \frac{1}{4}+ \Op(1/g) ,
\end{equation}
and for twist-two operators, $L=2$, we find
\begin{eqnarray}\label{eq:B2 LO}
 B_2(g)&=& (-\EulerGamma-\log g)f(g)- 4 g (1-\log 2) \nonumber\\
       &&-\big(1- \frac{6 \log2}{\pi} + \frac{3 (\log 2)^2}{\pi}\big)+\mathcal{O}(1/g)\,.	
\end{eqnarray}
The virtual correction thus cancels the $\EulerGamma$ dependence of \eqref{scaling} and 
with the explicit first two orders strong coupling scaling function $f(g)=4g-3\log 2 /\pi $ and our choice
of $g=\sqrt{\lambda}/4\pi$ we predict the string energy up to one-loop
\begin{eqnarray}
E-S&=&L+\gamma_L\big(\frac{\sqrt{\lambda}}{4\pi},S\big)\Big|_{L=2}=\big(\frac{\sqrt\lambda}{\pi}-\frac{3\log 2}{\pi }\big)\log\frac{4\pi\,S}{\sqrt{\lambda}}\nonumber\\
&+&\frac{\sqrt\lambda}{\pi}(\log 2-1)+1+\frac{6\log2}{\pi}-\frac{3(\log 2)^2}{\pi} \,,
\end{eqnarray}
and determine the constant $c$ of \cite{Beccaria:2008tg} to be 
\begin{equation}
 c=6 \log 2  + \pi \,.
\end{equation}
in agreement with \cite{Gromov}. With the result for $B_2$ given in \eqref{eq:B2 LO}, $B_L$ follows straightforwardly 
from \eqref{gamma1 final} for general, finite values of $L$.

With the all-loop quantization condition of \cite{Basso:2007wd} we can determine
subleading strong coupling corrections to the virtual correction $B_2(g)$ by solving
for the $c_p^\pm$ and accordingly performing the sums in \eqref{gamma1 final}. For 
the first terms we find 
\begin{eqnarray}
&& B_2(g+ c_1)=(\log \tfrac{2}{g}-\EulerGamma)f(g+c_1) -4g-1\nonumber\\
&&\phantom{B} + \frac{1}{g}\frac{\mbox{K}}{2\pi^2}-\frac{1}{g^2}\frac{9\zeta(3)}{2^8\pi^3}+\frac{1}{g^3}\Big(\frac{9\,\beta(4)}{2^7\pi^4}-\frac{\mbox{K}^2}{2^7\pi^4}\Big)\nonumber\\
&&\phantom{B} - \frac{1}{g^4}\Big(\frac{6831\,\zeta(5)}{2^{18}\pi^5}-\frac{423\,\mbox{K}\,\zeta(3)}{2^{13}\pi^5}\Big)+\,\mathcal{O}(1/g^5) ,
 \end{eqnarray}
where $c_1=\frac{3\log 2}{4\pi}$.
We made the same shift of the coupling constant as was suggested in \cite{Basso:2007wd}.
\subsection{Subleading spin corrections}
The NLIE is written in terms of the roots of the transfer matrix, the so called 
holes, see \cite{Belitsky:2006en, Freyhult:2007pz} for details. For $L=2$, 
using the explicit expression for the holes \cite{Belitsky:2006wg},
\begin{eqnarray}
u_h= &\pm&  \frac{2g^2+q_2}{\sqrt{-2\,q_2}},\nonumber\\
q_2=-\frac{1}{4} \Big( \big( 2M+2 &+& \gamma_2(g) \big) \big( 2M+\gamma_2(g) \big)+2 \Big),
\end{eqnarray}
it is possible to further expand the NLIE of \cite{Freyhult:2007pz}. Expanding to 
$\mathcal{O}((\log M)^2/M^2)$ changes the factor that multiplies $K(2gt,0)$ in 
\eqref{eq: complete density} to  
\begin{equation}
 \log M +\EulerGamma+\frac{f(g)}{2}\frac{\log M +\EulerGamma}{M}+\frac{1+B_2(g)}{2\,M},
\end{equation}
apart from this the integral equation remains unchanged.
In analogy with the above computation we find, for twist-two operators,
\begin{eqnarray}
 \gamma_2(g,M)&=&f(g) \Big(\log M+\EulerGamma +\frac{f(g)}{2}\frac{\log M + \EulerGamma}{M} \nonumber\\
 &&+ \frac{1+B_2(g)}{2\,M}\Big) +B_2(g)+ \ldots \,,
\end{eqnarray}
in perfect agreement with the result from string theory \cite{Beccaria:2008tg} and in 
agreement with reciprocity relations \cite{Basso:2006nk}. It is possible to formally continue 
the expansion but the next correction $\mathcal{O}((\log M)^2/M^2)$ receives contributions 
from wrapping effects \cite{Bajnok:2008qj}, and the NLIE therefore does not produce the correct result.
%
\section{Conclusion}
We have computed the finite order correction to the logarithmic scaling of large
spin operators of arbitrary twist at weak and strong coupling. As we obtained the integral 
equation from the infinite-volume NLIE we have unequivocally shown that the 
asymptotic Bethe ansatz \cite{Staudacher:2004tk} is capable to determine the first subleading 
term in the large spin expansion of twist operators, unaffected by wrapping effects. 

At strong coupling and twist $L=2$ we can reproduce the string theory results of \cite{Beccaria:2008tg} and 
determine the constant that arises in the one-loop sigma-model calculation. For this special value of twist
we also determined subleading correction in spin $M$ up to wrapping order in agreement with string theory.  

We have also established the first steps in order to fill the gap between certain Wilson line expectation
values and logarithms of multi-loop gluon scattering amplitudes as the virtual part $B$ is responsible for the difference
between the subleading singularities in these two quantities. It remains a challenge 
to ascertain an operator interpretation of the missing link to the complete collinear anomalous dimension, 
denoted as $G_{\mathrm{eik}}$, see \cite{Dixon:2008gr}. 
\vspace*{-2.5mm}
\begin{acknowledgements}
We have benefited from discussions with Lance Dixon, Adam Rej, Matthias Staudacher and Valentina Forini.
We are most grateful to Benjamin Basso and Gregory Korchemsky for valuable comments and sharing their insights into the 
all-loop quantization condition with us. This research was supported in part by the National Science 
Foundation under Grant No. PHY05-51164.
\end{acknowledgements}


\end{document}